# CURVATURE, PHASE SPACE, HOLOGRAPHY AND BLACK HOLE ENTROPY[*]

C Sivaram and Kenath Arun

Indian Institute of Astrophysics, Bangalore

**Summary:** This paper discusses the thermodynamics of a black hole with respect to Hawking radiation and the entropy. We look at a unified picture of black hole entropy and curvature and how this can lead to the usual black hole luminosity due to Hawking radiation. It is also shown how the volume inside the horizon, apart from surface area (hologram!), can play a role in understanding the Hawking flux. In addition holography also implies a phase space associated with the interior volume and this happens to be just a quantum of phase space, filled with just one photon. Generalised uncertainty principle can be incorporated in this analysis. These results hold for all black hole masses in any dimensions.

---

[*] This paper received the Honourable Mention for 2009 Awards for Essays on Gravitation

The study of black hole thermodynamics especially in connection with Hawking radiation and the vast entropy increase, involves a deep understanding of gravity, quantum physics, geometry, etc. An intriguing aspect is that black hole entropy is proportional to the area of the horizon [1] and given as: [2]

$$S \approx k_B \frac{GM^2}{\hbar c} \approx \frac{k_B c^3 A}{G\hbar} \qquad \ldots (1)$$

M is the black hole mass, A the horizon area, $k_B$, G and $\hbar$ are the Boltzmann constant, Newtonian gravitational constant and the Planck constant. Equation (1) can also be written as:

$$S \approx k_B \frac{A}{L_{Pl}^2} \qquad \ldots (2)$$

Where $L_{Pl}$ is the Planck length. $L_{Pl}^2 = \frac{\hbar G}{c^3}$, that is S can be expressed as $k_B$ times the ratio of the horizon area to the 'Planck area'.

The scaling of S with area A, rather than with the volume (like in conventional thermodynamic systems) led to the important suggestion of the holographic principle (i.e. all degrees of freedom contributing to the entropy reside on the surface like a hologram). This leads to the holographic bound [3, 4]:

$$S \leq \frac{k_B c^3 A}{4G\hbar} \qquad \ldots (3)$$

In recent papers [5, 6, 7], in trying to understand the entropy of dark energy (associated with a cosmological constant) and that of black holes in a cosmological context it was noted that the entropy of a space of constant curvature K can be written as:

$$S \approx \frac{k_B c^3}{G\hbar A K} \qquad \ldots (4)$$

Or
$$S \approx \frac{k_B}{L_{Pl}^2 K} \qquad \ldots (5)$$

(An observer in a curved space will measure a black body temperature $T \sim \frac{\hbar c}{k_B}\sqrt{K}$).

In fact equations (4) and (5) imply the upper entropy bound for a space of constant curvature K. [8]

For a black hole of mass M, we can replace K in equations (4) and (5), with the Schwarzschild curvature of the horizon $K_H \approx c^4/G^2M^2$ giving the (black hole) entropy as:

$$S \approx \frac{k_B}{L_{Pl}^2 K_H} \qquad \ldots (6)$$

formally identical to equation (2), $A \sim \frac{1}{K_H}$. Thus the black hole entropy is inversely proportional to the curvature of the horizon surface.

If the black hole 'grows' by accreting matter or there is merger of holes, the area increases but the curvature decreases, so the entropy increases. In the case of isolated black holes evaporating by Hawking radiation [9], there is decrease in horizon area and consequent increase in curvature and thus decrease in entropy.

The black hole shrinks owing to evaporation, the horizon area decreases till it reaches $L_{Pl}^2$ (i.e. the curvature increases to a maximum value of $K_{Pl} = \frac{1}{L_{Pl}} (\approx 10^{66} cm^{-2})$, when by equation (6) the entropy reaches the minimum $k_B$) (in thermodynamics $k_B$ is analogous to $\hbar$, the quantum of action). So equation (6) would suggest that black holes would evaporate leaving a remnant of Planck area (which has a maximal curvature) and the lowest unit of entropy, i.e. just $k_B$.

This is also consistent with superstring picture, where there is a minimal length (string length $\sim L_{Pl}$) or loop quantum gravity where area is quantised in units of $L_{Pl}^2$. So an outside observer just associates an entropy to the black hole which is inversely proportional to the horizon curvature. Evaporation without a remnant would imply an infinite curvature (singularity) which is untenable with this hypothesis which would hold for any curved space with a horizon (either a black hole or a cosmological space time such as that of de Sitter space time).

This would give a somewhat unified picture [8, 9, 10] in the sense that the product of the entropy and curvature is constant. Thus:

$$SK_H = \text{constant} = \frac{k_B}{L_{Pl}^2} \approx 10^{50} (cgs\ units) \qquad \ldots (7)$$

This also leads to other intriguing relations, involving the black hole luminosity due to Hawking radiation $(L_H)$.

We have:
$$L_H A_H = \text{constant} \approx \hbar c^2 \quad \ldots (8)$$

and also:
$$SL_H = \text{constant} \approx \frac{c^5}{G} k_B \approx 5 \times 10^{43} (cgs\ units) \quad \ldots (9)$$

Equation (9) would imply that when the black hole evaporates down to the Planck length $L_{Pl}$, its Hawking luminosity would be the maximum possible, as given by general relativity, i.e. $c^5/G \approx 3 \times 10^{52} W$. [11, 12]

If we consider Hawking radiation as a 'diffusion' process with a time scale $(t_H \sim G^2 M^3/\hbar c^4)$ [13], we can have an analogy with luminosity of stars, where the luminosity L is related to the photon diffusion time $\tau$ as:

$$L_S \approx \frac{Nk_B T}{\tau} \approx \frac{Nk_B T \lambda c}{R^2} \quad \ldots (10)$$

(T is the temperature, N the total number of nucleons, $\lambda$ is the photon mean free path, $\lambda \sim \frac{1}{n\sigma_T}$, $\sigma_T$ is the Thomson cross section, n the particle density, R is the radius of the star). For given N, T, equation (10) also implies:

$$L_S A_S = \text{constant} = Nk_B T \lambda c \quad \ldots (11)$$

$A_S$ is the stellar surface area. For a black hole, total energy $\approx Mc^2$, so that

$$L_H \approx \frac{Mc^3}{t_H} \approx \frac{\hbar c^6}{G^2 M^2} \quad \ldots (12)$$

Which is just the rate of emission of Hawking radiation.

For a star the temperature $T \approx \frac{GMm_P}{k_B R}$, ($m_P$ the proton mass), but for a black hole it is just corresponding to the largest wavenumber, $k_{max}$ (or smallest wavelength) which an external observer could measure (a quantum limit!), i.e.

$$T \approx \frac{\hbar c}{k_B R_S} \approx \frac{\hbar c^3}{k^B GM} \quad \ldots (13)$$

$(k_{max} \approx c^2/GM)$

Equation (13) is indeed the Hawking temperature. This suggests an interpretation of black hole entropy as an entropy of zero point modes [14, 15], i.e. analogous to the Casimir entropy for two parallel plates separated by a distance a, being given by the number of squares of edge a required to cover the plate area (i.e. $A/a^2$), the black hole entropy is the number of Planck length squares required to cover the horizon (i.e. $A_H/L_{Pl}^2$). Indeed in the formula for the energy density of zero point modes we have:

$$\varepsilon \approx \hbar c \int_0^{k_{max}} k^3 dk \approx \hbar c k_{max}^4 \qquad \ldots (14)$$

Substitute $k_{max} \approx c^2/GM$, and multiply by $c/4$ to get the Hawking flux. This times the horizon area, again just gives the rate of emission of Hawking radiation, $L_H$ as given by equation (12).

Equation (14) invoked the energy density. Does this imply any role for the volume (of the black hole) rather than the surface area (hologram!) in understanding the Hawking flux and entropy. In an earlier work [15], we had explicitly stated the subtle fact (almost elevated to the status of a theorem!) that for any black hole, Hawking radiation implies (at all times) the presence of just one photon (of wavelength $R_H$) in the entire volume $(\sim R_H^3)$ within the horizon. (This would also exactly imply the energy density as given in equation (14)). This can be converted into a surface integral by Gauss theorem, implying a quanta emission rate of $\approx c^3/GM$ through the horizon surface, which with the typical quanta energy given by equation (13) is just the Hawking emission rate.

So the presence of just one photon in the entire Schwarzschild volume (this is independent of the black hole mass or dimensions!) leads to the usual Hawking emission rate. It also leads to an evaporation time, proportional to the cube of the hole mass, i.e. $M^3$, i.e. the Schwarzschild volume. (i.e. horizon volume × time = constant $\approx c^2/\hbar$). This suggests a quantum diffusion process [16]. The presence of just one photon in a Schwarzschild volume is reminiscent of one particle per unit phase space.

Indeed taking the momentum corresponding to the $k_{max}$ (as given by the horizon) in equation (14), we can evaluate the phase space associated with the entire horizon interior. This surprisingly gives:

$$\left(d^3p\, d^3x\right)_{\text{inside horizon}} = \left(\frac{\hbar c^2}{GM}\right)^3 \times \left(\frac{GM}{c^2}\right)^3 = \hbar^3 \qquad \ldots (15)$$

Thus Hawking emission implies that the phase space associated with the black hole interior is just $\hbar^3$, i.e. the quantum of phase space. This relation holds for black holes of any mass (and in arbitrary dimensions $n > 3$). It also holds at any stage of evaporation. Thus holography implies phase space associated with the interior volume in just $\hbar^3$ and is filled with just one photon (or quantum).

For an observer falling into the black hole and crossing the horizon, again the phase space is conserved with some very interesting and important implications for formation of the singularity which is smeared by the quantisation of phase space [recent works in 17, 18]. All the above results hold for higher dimensions.

The consequence of incorporating the phase space constraint due to the Generalised Uncertainty Principle (GUP) can easily be explored. The phase space is now modified to [20]:

$$d^3x\, d^3p \geq \hbar^3 \left(1 + \lambda p^2\right)^3 \qquad \ldots (16)$$

where
$$\lambda \sim (l_S)^2 \sim (L_{Pl})^2 \qquad \ldots (17)$$

For macroscopic black holes, the corrections are quite small as $p \ll \frac{\hbar}{\sqrt{\lambda}}$. However as the hole continues to evaporate (curvature increases!) or for primordial black holes formed at the earliest epochs, $\lambda p^2 \approx 1$, so we have a factor of eight. Decay times would also be modified as phase space densities change [20].

This leads to an alternate view for quantum gravity where rather than space being quantised, we have phase space being quantised and treat this as the extended Einstein (geometric quantity) tensor, there being no source. This is being explored. [19]